\documentclass{elsart}
\usepackage{epsfig,graphicx, amssymb}
\begin{document}
\runauthor{Litak et al.}

\begin{frontmatter}
\title{Suppression of  chaos   by weak resonant excitations in a nonlinear 
oscillator with a non-symmetric potential}   
\author[Lublin1]{Grzegorz Litak\thanksref{E-mail},}
\author[Lublin2]{Arkadiusz Syta,}
\author[Lublin1]{Marek Borowiec}

\address[Lublin1]{Department of Applied Mechanics, Technical University of
Lublin,
Nadbystrzycka 36, PL-20-618 Lublin, Poland}
\address[Lublin2]{Department of Applied Mathematics, Technical University 
of
Lublin,
Nadbystrzycka 36, PL-20-618 Lublin, Poland}

\thanks[E-mail]{Fax: +48-815250808; E-mail:
g.litak@pollub.pl (G. Litak), \\ a.syta@pollub.pl (A. Syta), 
m.borowiec@pollub.pl (M. Borowiec)}

\begin{abstract}
We examine the Mielnikov criterion for transition to chaos in case of 
one degree of freedom nonlinear oscillator with non symmetric potential. 
This system subjected to an 
external periodic force shows homoclinic transition from regular 
vibrations to chaos just before 
escape from the potential well.
Especially we study the effect of a second resonant excitation with 
different 
phase on the system transition to chaos and propose a way of its 
control.
\end{abstract}

\end{frontmatter}

\section{Introduction}
In this note shell examine vibration and show possible   
non feedback control of  chaos in a simple, one degree of freedom, system 
subjected 
to external excitation
with a non-symmetric
stiffness given by the following   equation:
\begin{equation}
\label{eq1}
\ddot{x} + \alpha  \dot{x}
+ \delta x +\gamma x^2=\mu \cos{ \omega t}
\end{equation}
where $x$ is displacement  $\alpha \dot{x}$ is linear damping,
$\mu \cos{\omega t}$ is an external excitation while $\delta x$ and 
$\gamma x^2$
are  linear and  quadratic force terms:
\begin{equation}
F(x)=-\delta x - \gamma x^2.
\label{eq2}
\end{equation} 

Such systems, where quadratic term  brakes the symmetry of a potential $V(x)$ 
\begin{equation}
V(x) \neq V(-x),~~~~{\rm and}~~~~
F(x)=-\frac{ \partial V(x)}{\partial x},
\label{eq3}
\end{equation}
have been a subject 
of studies for many years 
\cite{Szabelski1985,Thompson1989,Szabelski1991,Szemplinska1993,Szemplinska1995,Rega1995,Litak1998,Rusinek2000,Rand1994,Rand2003}. 
These investigations were motivated by  possible 
applications in description of physical systems mostly mechanical 
\cite{Szabelski1985,Szabelski1991,Litak1998,Rusinek2000}
and electrical systems \cite{Szemplinska1993}. 
They were also linked to possible metastable states of atoms and 
appear in problems within the elastic theory 
\cite{Thompson1989,Thompson1984}.

Systems which show homoclinic orbits and can be tackled analytically
by perturbation methods. 
Namely by the Mielnikov method treating  \cite{Guckenheimer1983,Wiggins1990}
excitation and damping terms in higher order. 
 Such a treatment has been 
 performed to selected problems with both symmetric  and 
non-symmetric   nonlinear 
forces \cite{Guckenheimer1983,Wiggins1990,Rand2003} to derive the necessary condition for 
transition to chaotic motion.
On the other regular and chaotic regions of solutions in system parameters 
 can be stabilised by using an additional weak resonant excitation
\cite{Chacon2001,Cao2004a,Cao2004b,Litak2004}. 
In this note we shell apply this method  to
 the nonlinear system given by Eq. \ref{eq1}.
Combined 
with the Mielnikov approach it will predict analytically the range of
parameters which tam the chaotic behaviour.
The paper is divided into 5 sections.  After this introduction (Sec. 1)  
we perform Mielnikov analysis in Sec. 2. This discussion is followed 
by Sec. 3 where we include a week resonant excitation term. Its useful role in  system control is 
shown there. The 
analytic predictions are confirmed by means of numerical simulations 
(Sec. 4). We 
ending up in Sec. 5 with  conclusions.

\section{Melnikov analysis}
Note in this section we follow the discussion initiated by Thompson 
\cite{Thompson1989} where he derived the analytic formula for 
a critical amplitude of a nonlinear oscillator described by a similar (to 
our Eq. 1) 
equation. We decided to include this section as an important introductory 
part to our main results to be given in the next section. 
 
Thus we start our study from the second order equation of motion 
\ref{eq1}.
Transforming it into two differential equations of the first order we get
\begin{eqnarray}
&& \dot{x} = v \label{eq4}\\
&& \dot{v}  = -\delta x - \gamma x^2 + \epsilon \left[ -\tilde{\alpha}v + 
\tilde{\mu} \cos{\omega t}
\right]. \nonumber
\end{eqnarray}
Looking for stable and unstable manifolds we have introduced small 
parameter 
$\epsilon$ to the above equations and renormalised parameters $\tilde 
\alpha$ and $\tilde \mu$ via $\alpha = \epsilon \tilde 
\alpha$ and $\mu= \epsilon \tilde \mu$, respectively.

\begin{figure}[htb]
\hspace{0.5cm}
\epsfig{file=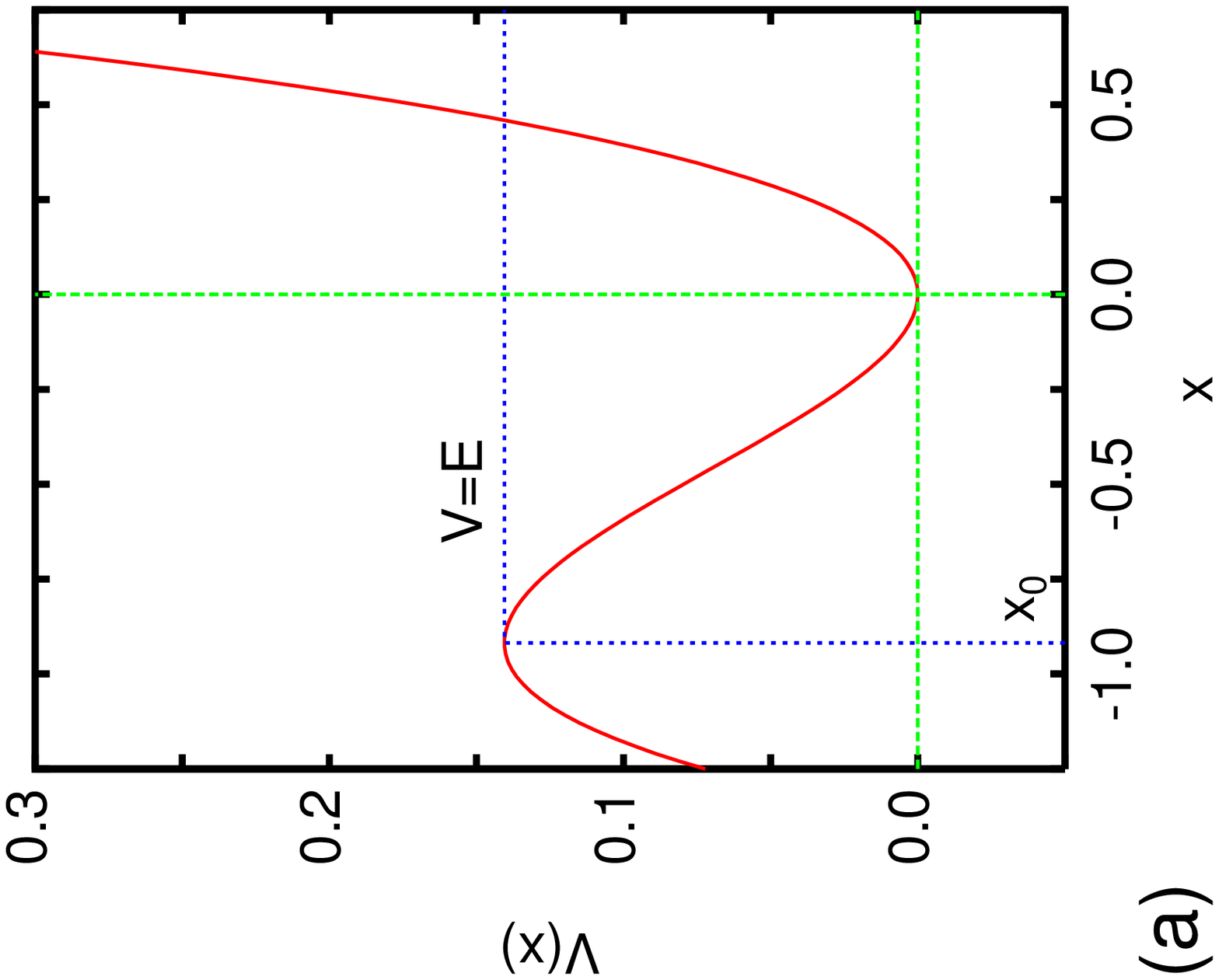,width=5.5cm,angle=-90}

\vspace{-4.4cm}
\hspace{6cm}
\epsfig{file=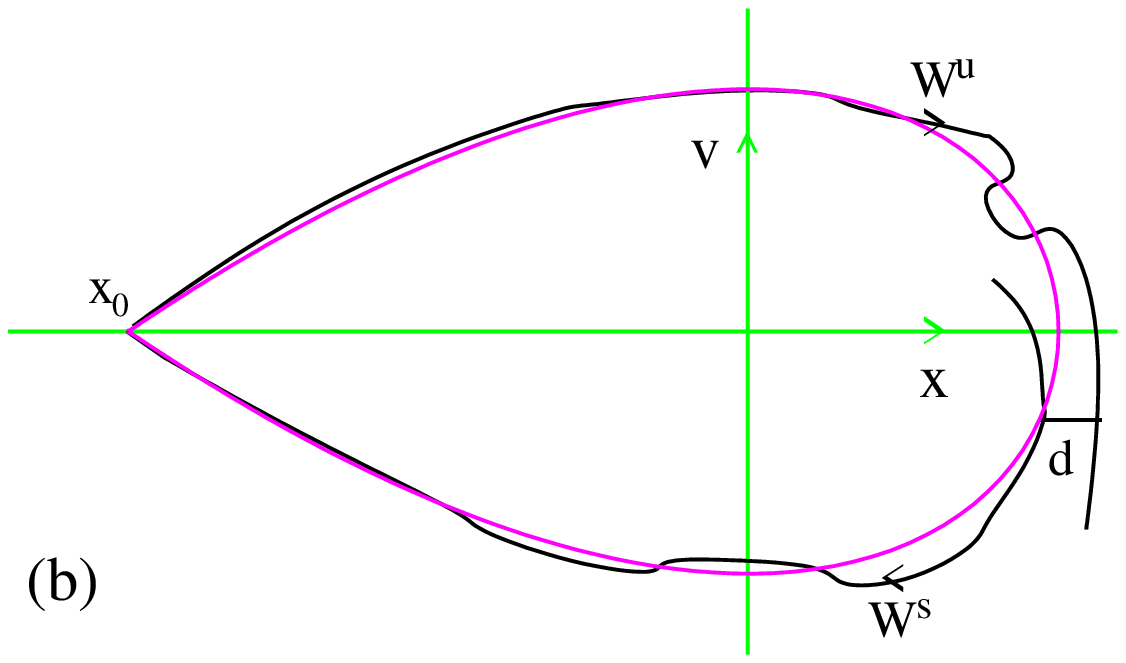,width=7.2cm,angle=0}
 \caption{ \label{fig1} (a) The potential well of the unperturbed  Hamiltonian (Eq.
\ref{eq4}) for
$\delta=1$ and
$\gamma=1.089$, the energy level $V=E$ corresponds to the Hamiltonian system with a tangent point at $x_0$  ; (b) stable $W^S$ and
unstable $W^U$
manifolds for unperturbed (gray) and a damped and
excited system (in black). The distance $d$ between $W^S$ and $W^U$ can be
described by Mielnikov function $M(t_0)$ (Eq. \ref{eq9}). Note $x_0$ indicates the
local
extremum of potential $V(x)$ (Fig. 1a) which simultaneously represents  a
saddle point in the phase plane (Fig. 1b).}
\end{figure}

The corresponding unperturbed Hamiltonian have the following form:
\begin{equation}
H^0= \frac{v^2}{2} + V(x), \label{eq5}
\end{equation}  
where 
\begin{equation}
V(x)=\frac{\delta x^2}{2} + \frac{\gamma x^3}{3} \label{eq6}
\end{equation}
is the potential of a non-symmetric well plotted in Fig. \ref{fig1}a.
This function has the local peak  at the point
\begin{equation}
x_0=-\frac{\delta}{\gamma}. \label{eq7}
\end{equation}
Existence of this point with a horizontal tangent make possible 
homoclinic bifurcations of the system i.e. potential transition from a regular to 
chaotic 
solution.
After simple integration  (Appendix A) we get homoclinic orbits 
(Fig. \ref{fig1}b) as:
\begin{eqnarray}
&& x^* =  \frac{\delta}{\gamma} \left( \frac{1}{2} - \frac{3}{2} \tanh^2 
\left( 
\frac{\sqrt{\delta} ( t-t_0) }{2} \right) \right),
\nonumber \\
&& v^* = 
- \frac{3}{2} \frac{ \delta \sqrt{\delta} ~\tanh \left(
\frac{\sqrt{\delta} ( t-t_0)}{2} \right)}{ \gamma ~\cosh^2\left(
\frac{\sqrt{\delta} ( t-t_0)}{2} \right) }. \label{eq8} 
\end{eqnarray}

Note the characteristic saddle point $x_0$ is going to be reached in 
exactly defined albeit infinite time 
$t$  corresponding to $+\infty$ and $-\infty$
for stable and unstable orbits, respectively.

\begin{figure}[htb]
\centerline{
\epsfig{file=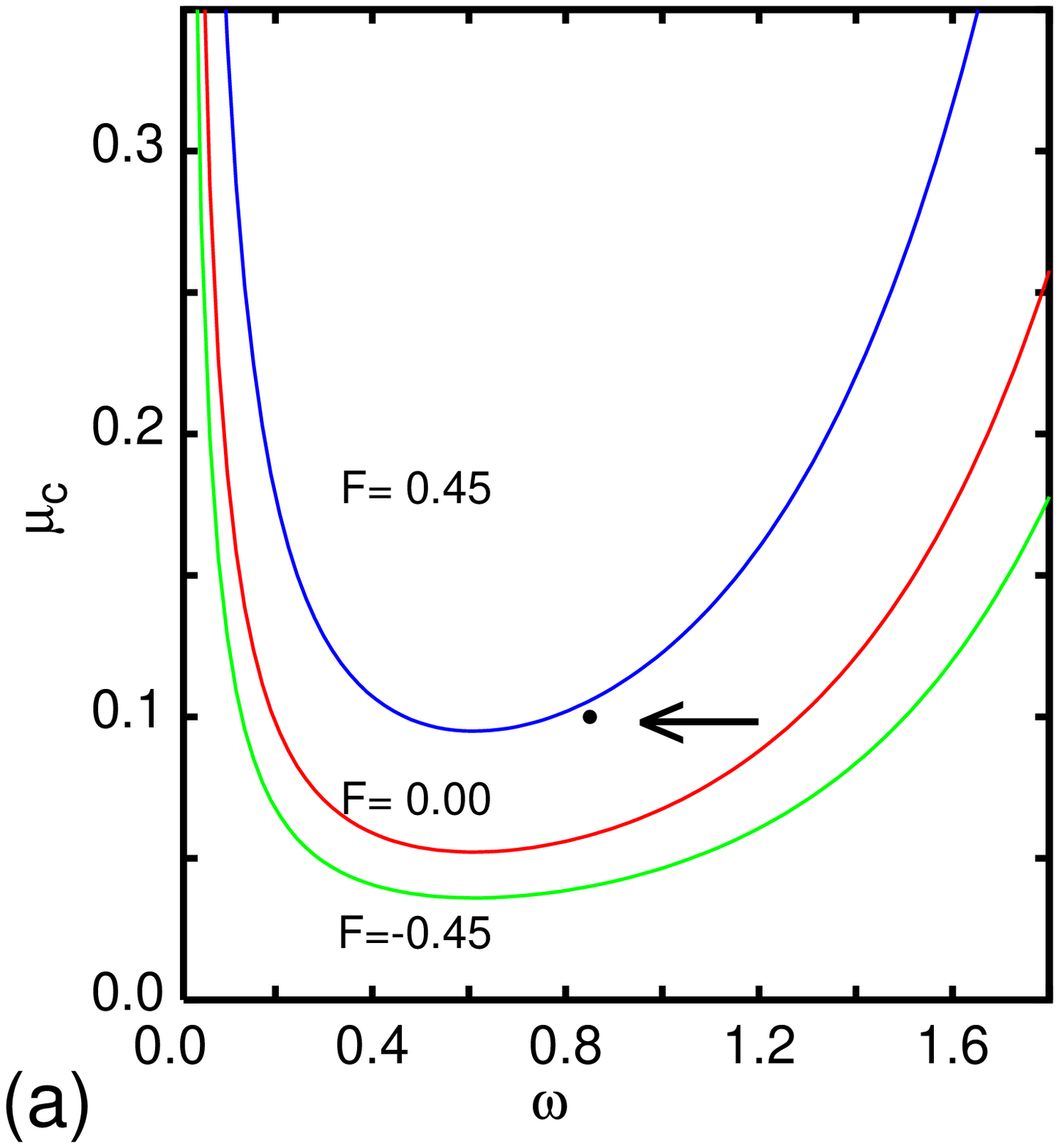,width=5.5cm,angle=-90}
\epsfig{file=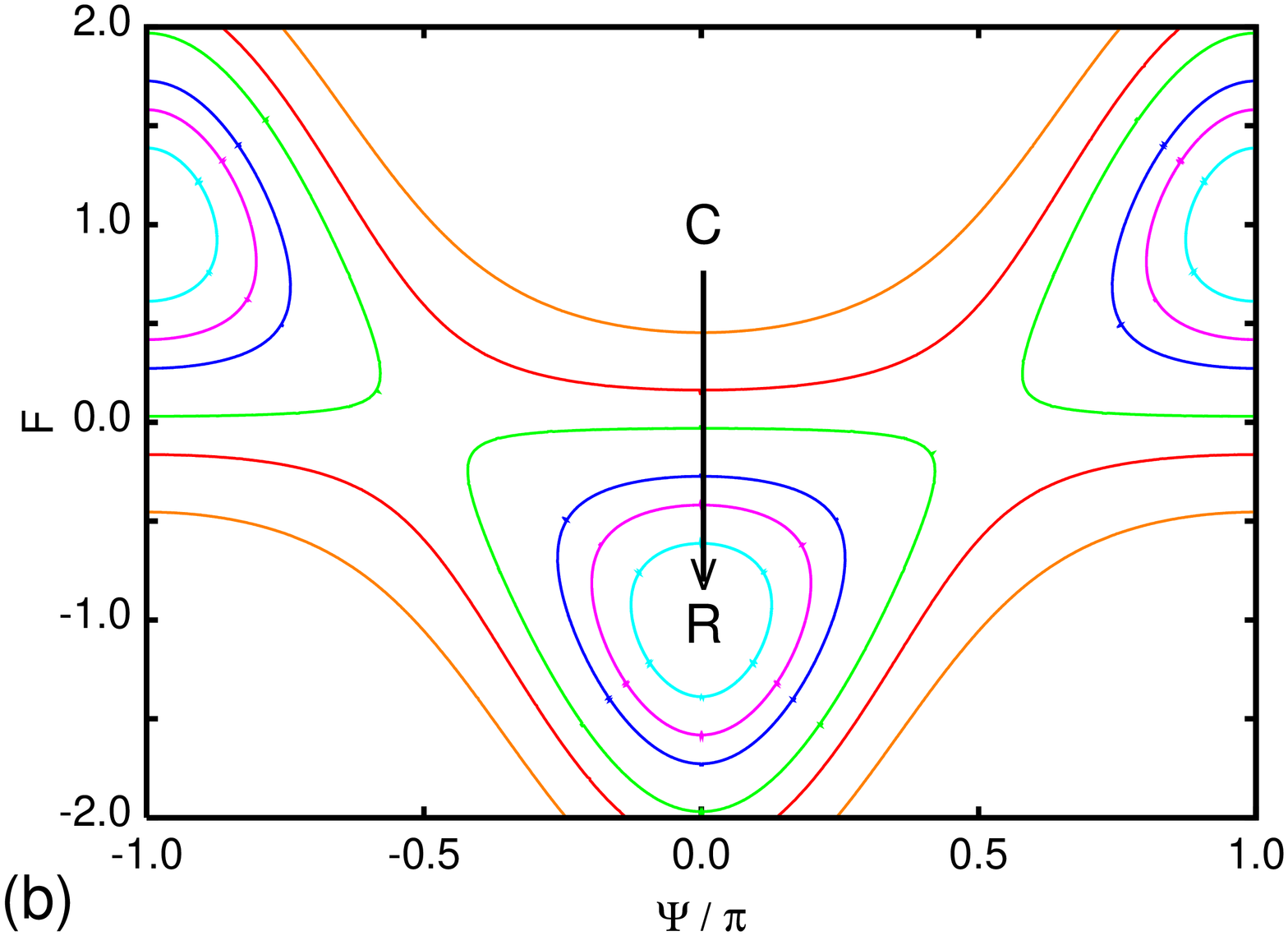,width=5.5cm,angle=-90}}
 \caption{ \label{fig2} (a) Critical value of $\mu_c$ plotted against $\omega$ for 
system parameters: $\gamma=1.089$, $\delta=1.0$. Control 
parameters were taken as $\Psi=\pi$ 
while $F=-0.45,~0.0$ and 0.45. The singular point below the upper curve corresponds to $\mu=0.1$ and $\omega=0.85$ used in 
numerical calculations - Figs. \ref{fig4}-\ref{fig5}.  (b) Phase diagram in a 
$\Psi$--$F$ plane for 
a number of $\mu$ values,
corresponding lines. The arrow indicate the direction of $\mu$ increase:
$\mu$ = 0.04, 0.05, 0.06, 0.08, 0.10 and 0.15, respectively.  Here 'R' and 'C' denotes regions with regular and chaotic solutions.  
}
\end{figure}

In case of perturbed orbits $W^S$ and $W^U$ the distance between them is given 
by 
the Mielnicov function $M(t_0)$:
\begin{equation}
M(t_0) = \int_{- \infty}^{ + \infty}  h( x^*, v^*)  \wedge g( x^*,
v^*) {\rm d} t \label{eq9}
\end{equation}
where the corresponding differential forms $h$ as the gradient of unperturbed 
Hamiltonian (Eq. \ref{eq3}) leading to equations of motion 
\begin{equation}
\frac{\partial H^0}{\partial x} = -\dot{v}, ~~~ \frac{\partial 
H^0}{\partial 
v} 
= \dot{x}, \label{eq10}
\end{equation}
while $g$ as its 
perturbation form of the above (Eq. \ref{eq4}):
\begin{eqnarray}
h &=& \left(\delta x + \gamma x^2\right) {\rm d} x  + v {\rm d}v, \\
g &=& \left( \tilde{ \mu} \cos{\omega \tau} - \tilde{\alpha} v  
\right)  {\rm d}x \label{eq11}
\nonumber
\end{eqnarray}
are
defined on homoclinic manifold $(x,v)=(x^*,v^*)$ (Eq.  \ref{eq6}, Fig. \ref{fig2}b).
From the above (Eqs. \ref{eq8}-\ref{eq11})
the Mielnikov integral is given by
\begin{equation}
M(t_0)=  \int_{- \infty}^{ + \infty} {\rm d}t~~ \left( \tilde \mu  v^* 
\cos{\left(\omega 
t \right)} - \tilde \alpha v^{*2} \right) \label{eq12}
\end{equation}
After substituting $x^*(t)$ and $v^*(t)$ by formulae given in Eq. 
\ref{eq8} and 
taking $\tau=\sqrt{\delta}t/2$
we get:

\begin{equation}
M(t_0) = -  \frac{9}{2} \tilde \alpha \frac{\delta^2}{\gamma^2} 
\sqrt{\delta} 
I_1 - 3 \tilde \mu  \frac{\delta}{\gamma} I_2, \label{eq13}
\end{equation}
where 

\begin{eqnarray}
I_1 &=&   \int_{- \infty}^{ + \infty} {\rm d}t~~ \frac{\tanh^2 
\tau}{\cosh^4 \tau} \nonumber \\
I_2 &=&  \int_{- \infty}^{ + \infty} {\rm d}t~~ \frac{\tanh \tau}{\cosh^2 
\tau} \cos{\frac{2 \omega \left(\tau +\tau_0\right)}{\sqrt{\delta}}} \label{eq14}
\end{eqnarray}
After evaluation of these elementary integrals we get 
condition of homoclinic transition to chaos, 
corresponding to a horseshoe type of 
cross-section and can be written as:
\begin{equation}
\bigvee_{ \displaystyle t_0}~~~~  M(t_0)=0,~~~~ \frac{\partial M(t_0)}{\partial t_0}  \neq 0.
\label{eq15}
\end{equation}
Evaluating above integrals (Eq. \ref{eq14})
after some lengthly algebra the last condition (Eq. \ref{eq15}) leads to  
a critical value of excitation amplitude 
$\mu_c$ for 
which
\begin{equation}
\mu_c= \frac{1}{5} \frac{\delta^2 \sqrt{\delta} \alpha}{ \pi \omega^2 \gamma} \sinh{ \left( \frac{\omega \pi}{\sqrt{\delta}} 
\right)}. \label{eq16} 
\end{equation}

As a result we get the critical amplitude $\mu_c$ versus frequency $\omega$, 
 which is plotted in Fig. \ref{fig2}(a) (the middle curve for  $F=0$). One should note here 
that the function has
a local minimum  at the point $\omega \approx 0.6$. Note that in spite of using 
$\cos(\omega t)$ in place  of $\sin(\omega t)$, in
the external excitation,
discussed by Thompson \cite{Thompson1989} in 
the external excitation  this result appears to be  the same.

\section{Effect of a weak resonant excitation}
Let us consider an additional excitation term in starting equation (Eq. \ref{eq1})
\begin{equation}
\label{eq17}
\ddot{x} + \alpha  \dot{x}
+ \delta x +\gamma x^2=\mu \cos{ \left(\omega t \right)}+\mu F \cos{\left( \omega t + \Psi \right)}, 
\end{equation} 
where $F$ is a scaling coefficient while  $\Psi$ represents a  phase of a weak resonant excitation.
Now instead of Eq. \ref{eq4} one can separate unperturbed and small perturbation parts in the following differential 
equations  of the first order:  
\begin{eqnarray}
&& \dot{x} = v \label{eq18}\\
&& \dot{v}  = -\delta x - \gamma x^2 + \epsilon \left[ -\tilde{\alpha}v + 
\tilde{\mu} \cos{\left( \omega t \right)} +\tilde{\mu} F \cos{\left( \omega t + \Psi \right)}
\right]. \nonumber
\end{eqnarray}
Now one can repeat most of calculations from the previous section assuming that
the excitation term is composed of two parts.
After simple algebra we get:
\begin{equation}
\cos{\left( \omega t \right)} +F \cos{\left( \omega t + \Psi \right)} = 
 \sqrt{1+F^2+2F\cos{\Psi}}~\cos{\left( \omega t + \alpha \right)}, \label{eq19}
\end{equation}
where
\begin{equation}
\alpha= \arctan{\left(
 \frac{1+F\cos{\Psi}}{F\sin{\Psi}}
\right)} \label{eq20}
\end{equation}
\begin{figure}[htb]
\centerline{
\epsfig{file=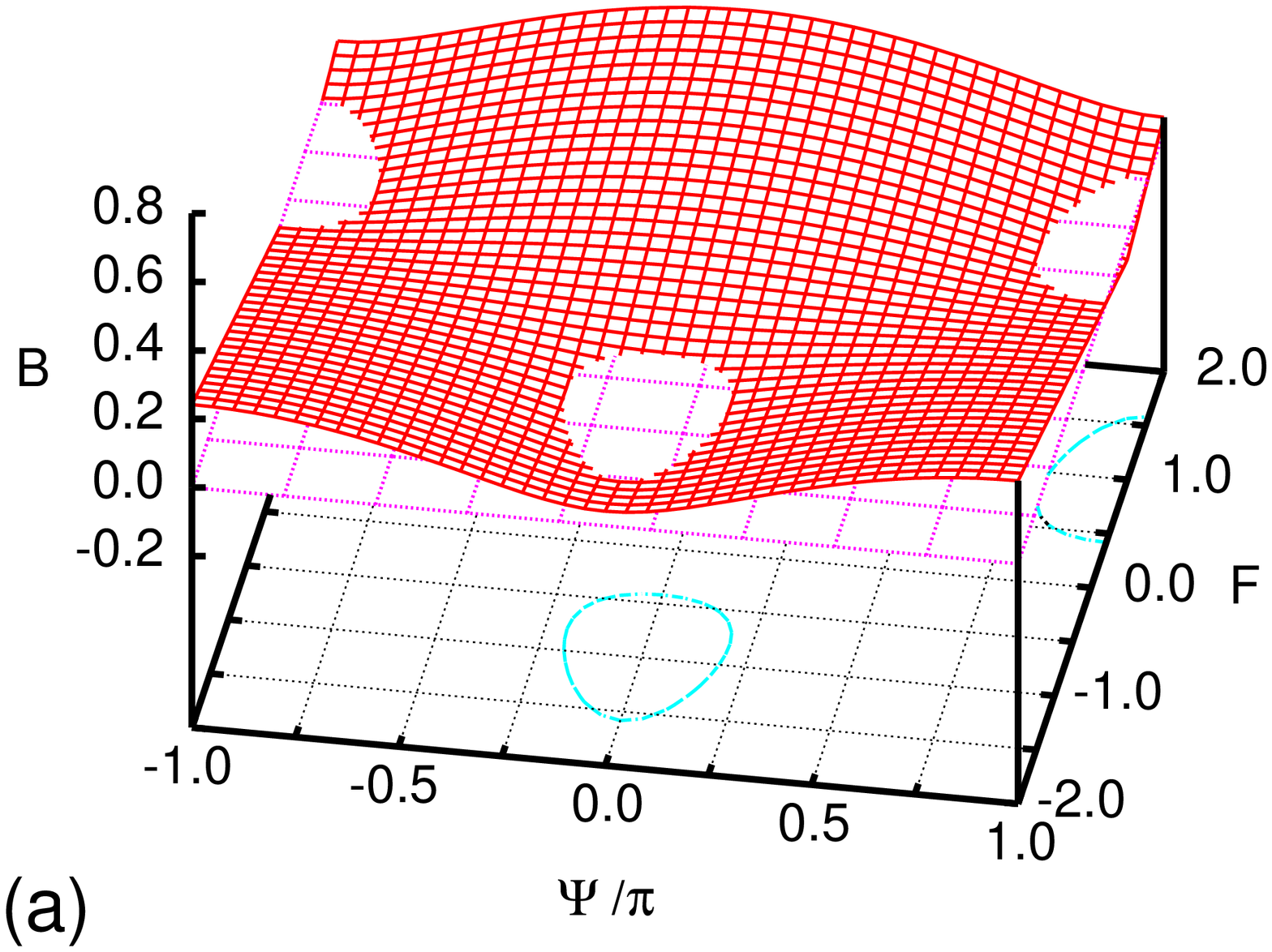,width=5.5cm,angle=-90}
\epsfig{file=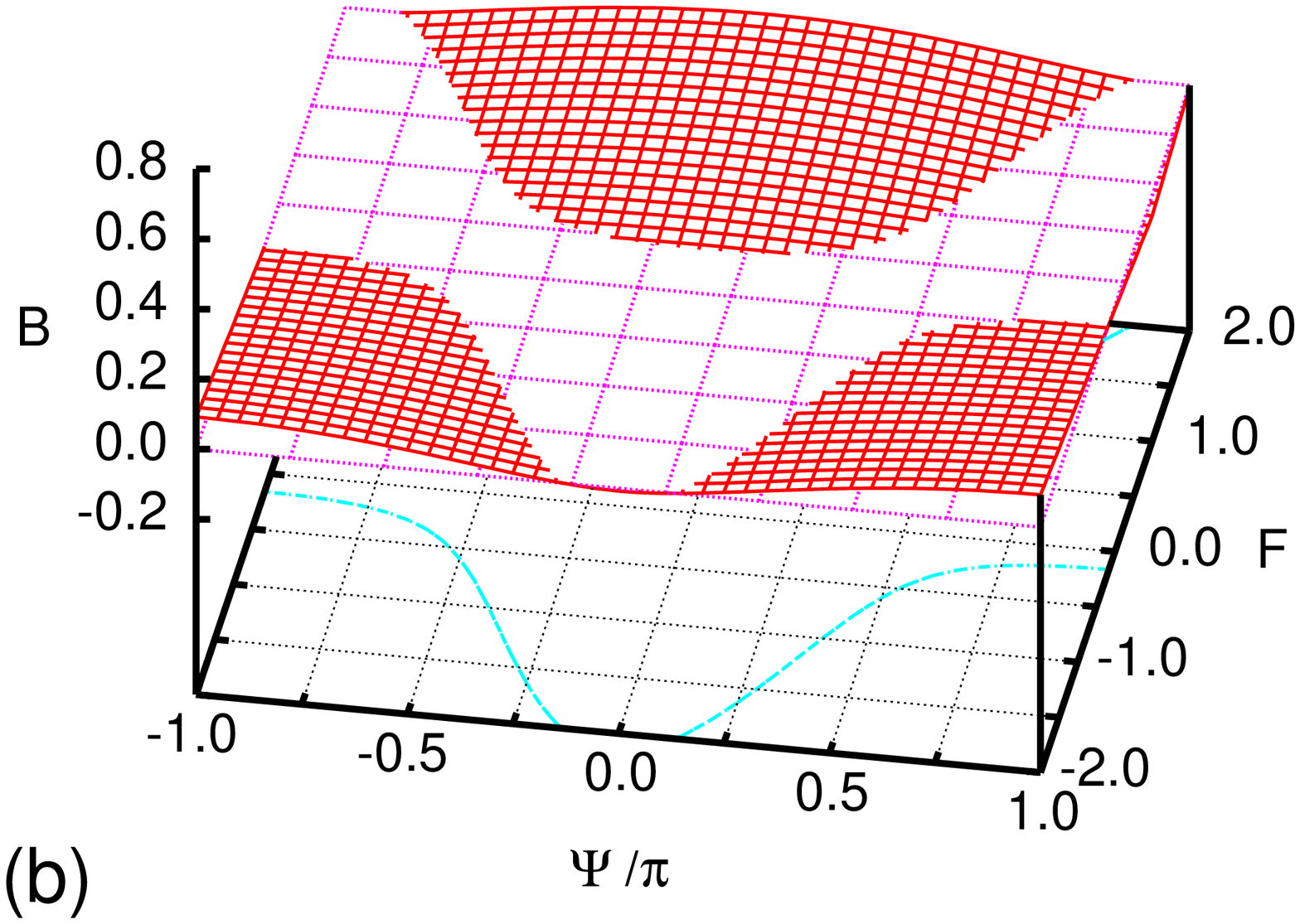,width=5.5cm,angle=-90}}
 \caption{ \label{fig3} Function $B(F,\Psi)$ for $\mu$ = 0.1 (a) and 0.05 (b)
respectively.
 }
\end{figure}

Therefore in this case the critical amplitude $\mu_c$ depends on
control parameters $F$ and $\Psi$:
\begin{equation}
\mu_c= \frac{1}{5} \frac{\delta^2 \sqrt{\delta} \alpha}{ \pi \omega^2 \gamma} \sinh{ \left( \frac{\omega \pi}{\sqrt{\delta}}
\right)} \frac{1}{\sqrt{1+F^2 +2F \cos \Psi}}. \label{eq21}
\end{equation}
This is our principal  result in this paper.
In Fig. \ref{fig2}a   we have plotted the results of $\mu_c$ for $\Psi= \pi$ and $F=-0.45$, 0.0, 0.45. Note that in case 
of $F=0$
the additional 
weak resonant excitation is absent but for any other cases it influences the system vibrations.
It can drive the system away ($F=0.45$), or into ($F=-0.45$)  
the regions with a
 potential chaotic solution.   To show the separation of regions with regular solutions and potential chaotic ones we calculated 
function $B(F,\Psi) \sim M(t_0)$ for a properly chosen $t_0$ (Eq. \ref{eq15}): 
In our case we defined $B$ as follows:
\begin{equation}
B =\frac{1}{5} \frac{\delta^2 \sqrt{\delta} \alpha}{ \pi \omega^2 \gamma} \sinh{ \left( \frac{\omega \pi}{\sqrt{\delta}}
\right)} - \mu~{\sqrt{1+F^2 +2F \cos \Psi}} \label{eq22}.
\end{equation}
In Fig. \ref{fig2}b we present the corresponding nodal lines for given amplitudes 
$\mu$ = 0.04, 0.05, 0.06, 0.08, 0.10 and 0.15, respectively.
On the other hand in Figs. \ref{fig3}a and b we show the $B$ as a function of $F$ and $\Psi$, for $\mu$ = 0.1 and 0.05 
(\ref{fig3}a 
and b 
respectively).
Note, the nodal lines separating regular region from the regions where chaotic solutions occur can be 
easily obtained  by 
cross-section 
with $B=0$ plane. 

\begin{figure}[htb]
\centerline{
\epsfig{file=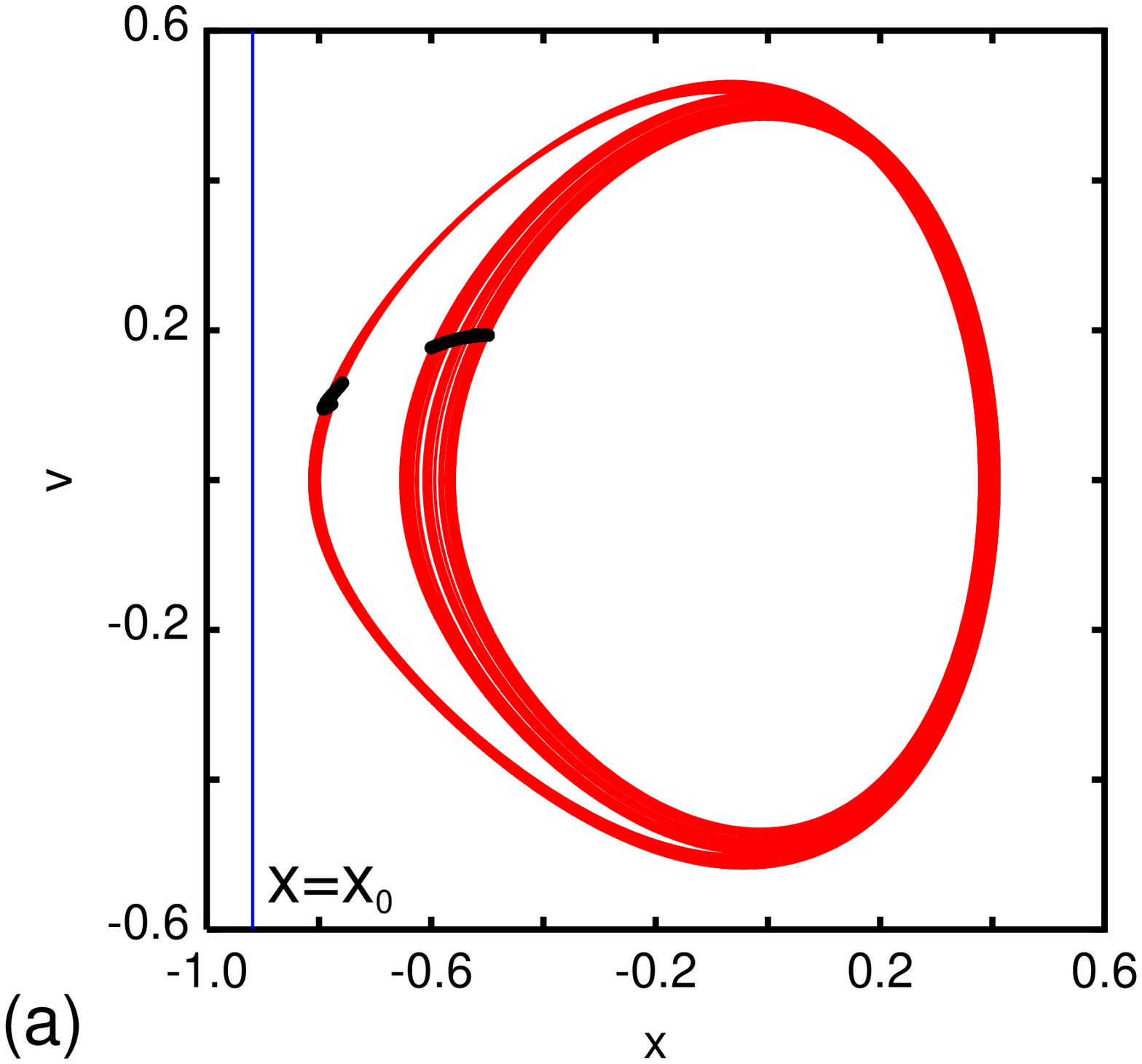,width=5.5cm,angle=-90}
\epsfig{file=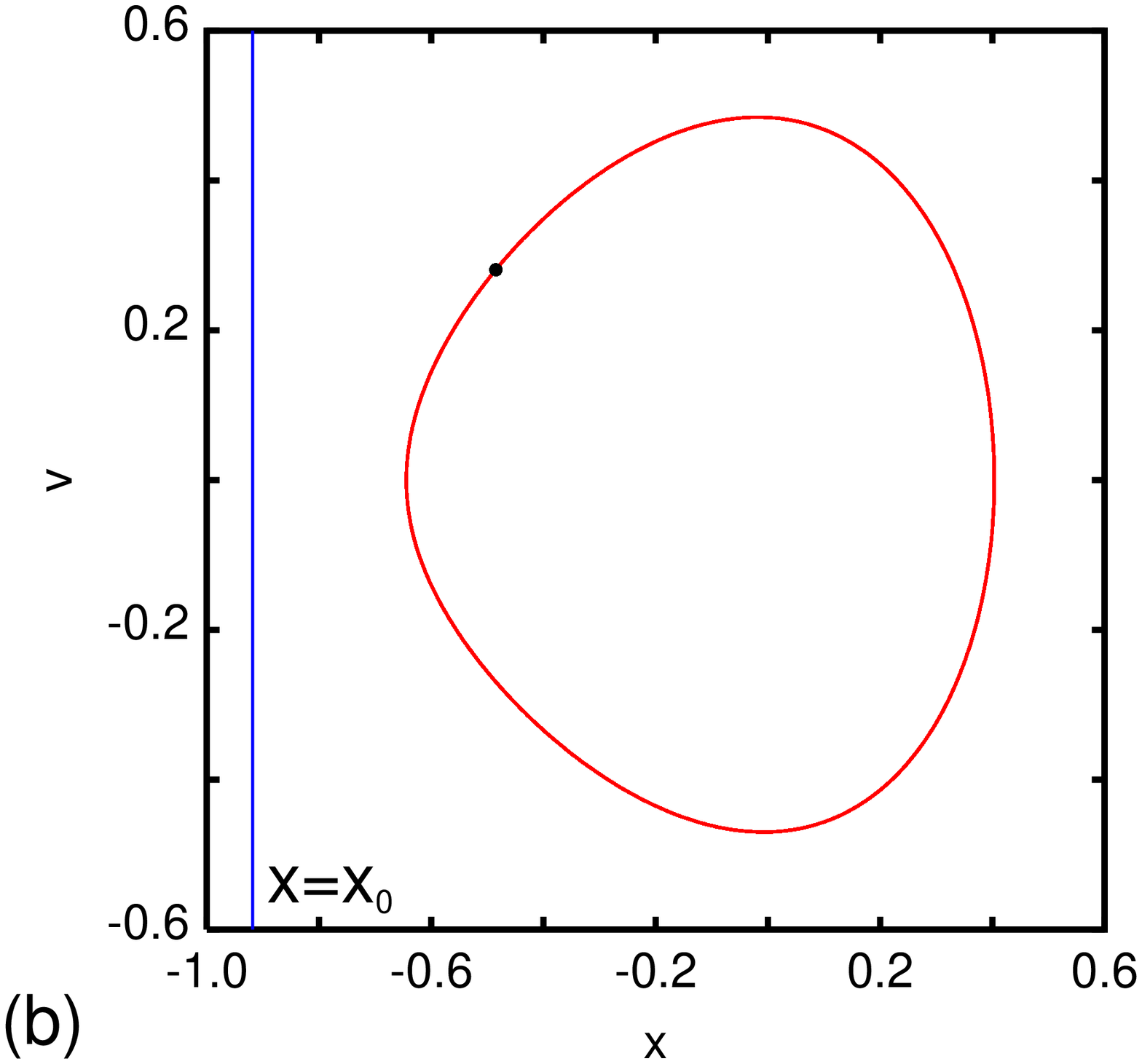,width=5.5cm,angle=-90}}
 \caption{ \label{fig4} Phase portraits plotted with lines together with Poincare sections denoted by points
for $F=0$ (a) and $F=0.45$ (b). $\mu=1$ and all other system parameter as in Fig. \ref{fig2}a. Note system parameters 
correspond
to
the point denoted with arrow in Fig. \ref{fig2}a. }
\end{figure}

\section{Numerical results}

\begin{figure}[htb]
\centerline{
\epsfig{file=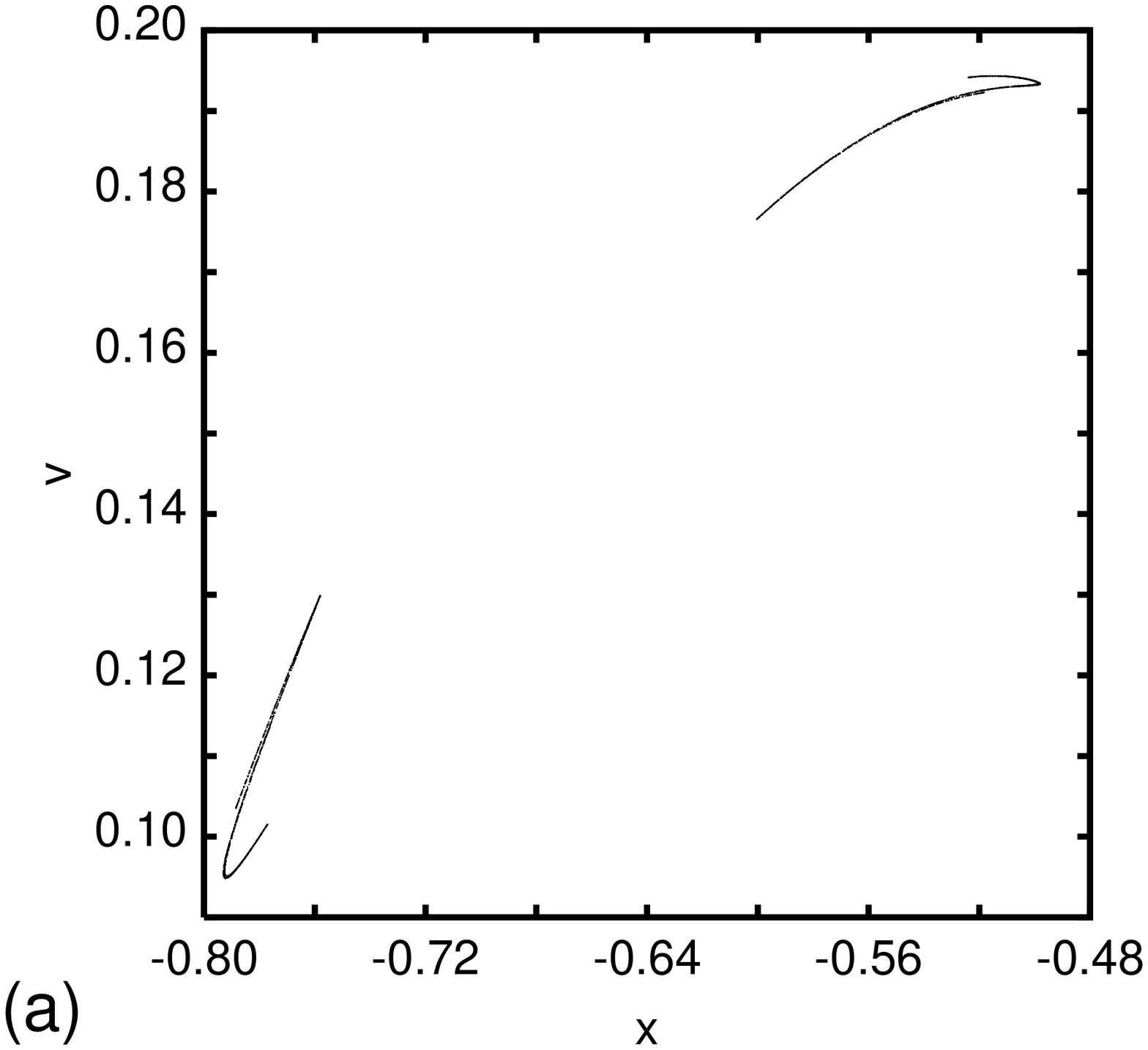,width=5.5cm,angle=-90}
\epsfig{file=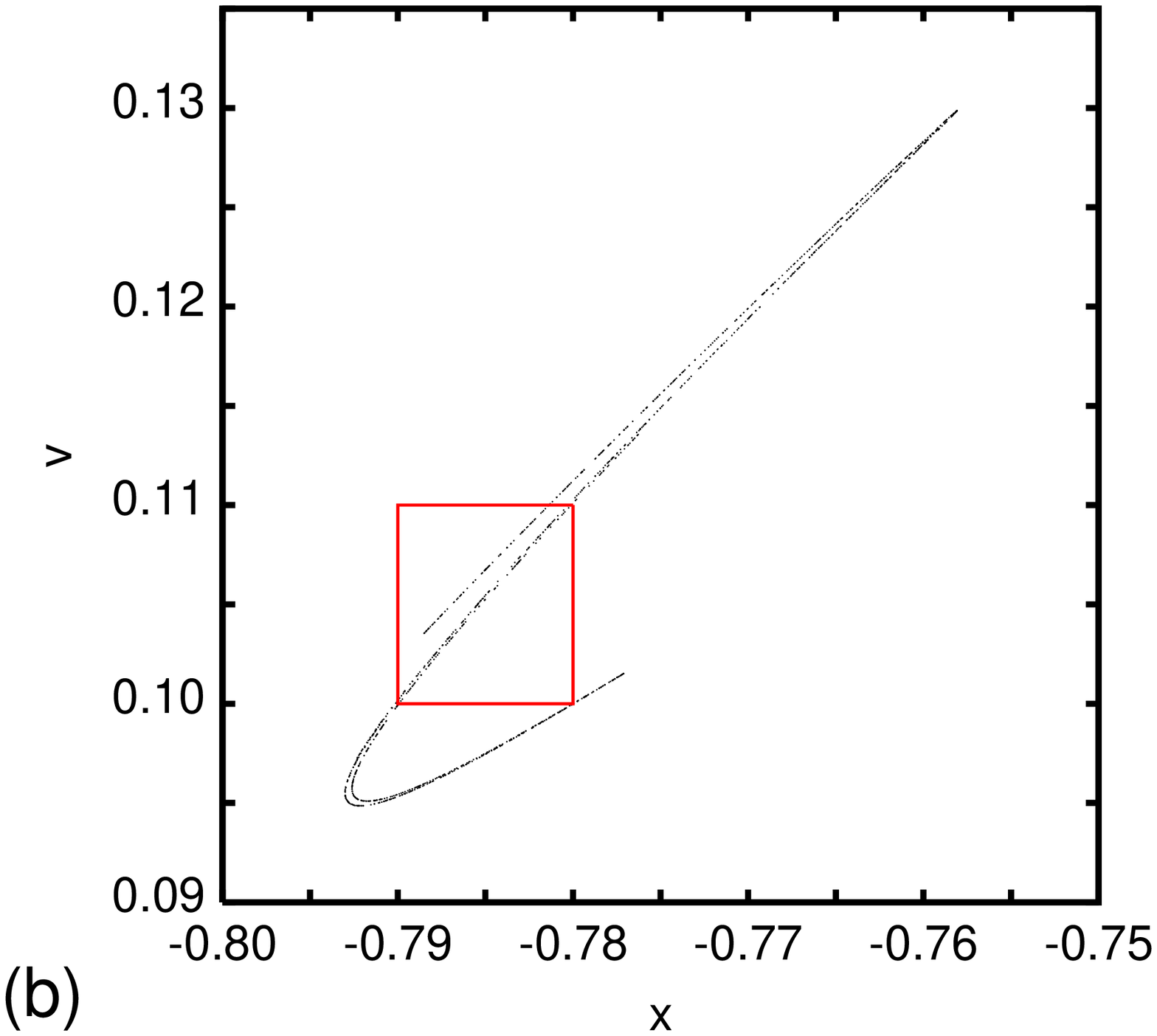,width=5.5cm,angle=-90}}

\centerline{
\epsfig{file=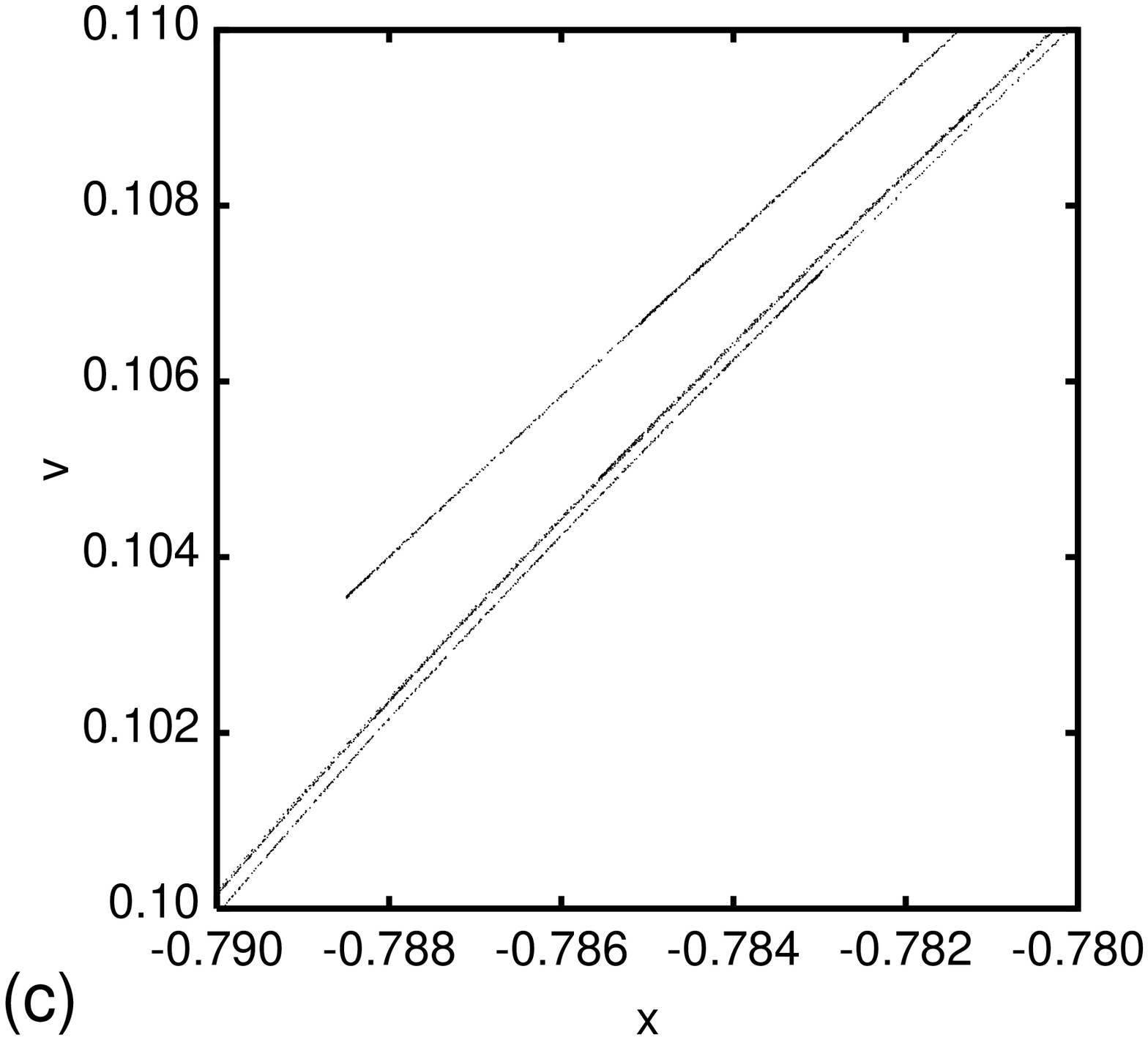,width=5.5cm,angle=-90}}

 \caption{ \label{fig5} (a) Chaotic attractor (magnified from Fig. \ref{fig4} 
and its fractal structure  (b-c).}
\end{figure}

To confirm the analytic predictions we have done numerical simulations for system parameters chosen as in  
point denoted with arrow in Fig. \ref{fig2}a. In Fig. \ref{fig4} we have plotted phase portraits and corresponding 
Poincare sections for $F=0$ (a),  $F=0.45$ (b) and $\Psi=\pi$. 
As expected from our initial discussion, based on approximate 
Mielnikov 
approach, we find chaotic solution in Fig. \ref{fig4}a. The maximal Lyapunov exponent 
appeared to be positive 
( $\lambda_1 \approx 0.05$. 
On the other hand Fig. \ref{fig4}b shows a regular type of motion synchronised with a periodic  excitation force. 
Note that the above numerical (Fig. \ref{fig4})  results confirms the analytic 
investigations (Fig. \ref{fig2}).  
To show the fractal structure the strange attractor (in Fig. \ref{fig4}a)
has been also magnified in Fig. \ref{fig5}. Note, different range of axes in Fig. 
\ref{fig5}a-c.
The characteristic double lines structure occur with the scaling 
factor which was estimated to be 4.

\section{Last remarks and conclusions}
Using the Mielnikov method we have got the analytical formula for
transition to chaos in a one degree of freedom, system subjected
to parametric excitation
with a non-symmetric
stiffness with self-excitation term. Including a weak resonant term we have been able to control
the type of solution. Namely it suppressed the chaotic solution by shrinking the lower bound lines of chaos in a  'phase' diagram   
(Fig. \ref{fig2}b).

In our numerical simulations of system vibration 
we have  used the same 
initial conditions $x_{ini}=x_0+ 0.1$ and $v_{ini}=0.1$ representing the point in the phase 
plane relatively close saddle one 
$(x,v)=(x_0,0)$.

Our basic chaotic solution were very close to that one obtained by 
Thompson \cite{Thompson1989} with ($|\gamma|=1.0$ and $\mu=0.1089$).
Although we have not discussed the problem of escape from the potential well 
 we changed 
slightly
his parameters   
($ \gamma \rightarrow 1.089$ and   $\mu \rightarrow 0.1$) purposely. This has helped us   
to have more robust chaotic 
solution away the 
escape point.

In the present analysis we have used only positively defined  $\gamma$ and $\delta$ (see Fig. 
\ref{fig1}a). Nevertheless  we claim that our results 
are general.
Note that another choice 
of 
$\gamma$ sign would only mirror our solution  along the $x$ axis ($x \rightarrow -x$). On the 
other hand changing $\delta
\rightarrow  -|\delta|$    
would also rescale  $x$ axis and simultaneously  shift the saddle point as well 
as the local minimum of the potential:
\begin{equation}
 V(x)= -\frac{|\delta|}{2}x + \frac{\gamma}{3}x^2 = \frac{|\delta|}{6}x' + 
\frac{\gamma}{3}(x')^2 -\frac{1}{6},  
\end{equation}
where 
$x'=x-|\delta|/\gamma$.    

\section*{Acknowledgements}
We would like to acknowledge partial support from Polish State Committee of Scientific 
Research.

\appendix{Appendix A}
\def\thesection{A}
\def\theequation{A.\arabic{equation}}  

Starting from unperturbed Hamiltonian $H^0$ (Eq. \ref{eq5}) we note that at the
saddle point
$x_0$ (Eq. \ref{eq7}) the system
velocity is zero $v=0$ and the energy has only
its
potential part
\begin{equation}
E=V(x=-\frac{\delta}{\gamma})=\frac{1}{6} \frac{\delta^3}{\gamma^2}.
\end{equation}
transforming Eq. \ref{eq5} for chosen constant energy (Eq. A1)
\begin{equation}
E=\frac{1}{6} \frac{\delta^3}{\gamma^2}={\rm const}
\end{equation}
we get the following expression for velocity:
\begin{equation}
v= \frac{{\rm d} x}{{\rm d} t} =\sqrt{2 \left(\frac{1}{6}
\frac{\delta^3}{\gamma^2}-
\frac{\delta x^2}{2} - \frac{\gamma
x^3}{3}\right)},
\end{equation}
Now one can perform integration  over $x$
\begin{equation}
t-t_0=\int {\rm d} x \frac{1}{\sqrt{2 \left(\frac{1}{6}
\frac{\delta^3}{\gamma^2}-
\frac{\delta x^2}{2} - \frac{\gamma
x^3}{3}\right)}},
\end{equation}
where $t_0$ represents an integration constant.
Finally, we get so called homoclinic orbits (Fig. 1b).
\begin{eqnarray}
&& x^* =  \frac{\delta}{\gamma} \left( \frac{1}{2} - \frac{3}{2} \tanh^2 
\left(
\frac{\sqrt{\delta} ( t-t_0) }{2} \right) \right),
\nonumber \\
&& v^* =
- \frac{3}{2} \frac{ \delta \sqrt{\delta} ~\tanh \left(
\frac{\sqrt{\delta} ( t-t_0)}{2} \right)}{ \gamma ~\cosh^2\left( 
\frac{\sqrt{\delta} ( t-t_0)}{2} \right) }. \label{eqA5}
\end{eqnarray}



\end{document}